\documentclass{article}
 
\usepackage[labelsep=period]{caption}
\usepackage{gensymb}
\usepackage{cite}
\usepackage{amsmath,tabularx,makecell}
\usepackage{graphicx}
\usepackage{array}
\usepackage{epstopdf}
\usepackage{caption}
\usepackage{float}
\usepackage{multirow}
\usepackage{authblk}
\usepackage{hyperref}
\usepackage{mathtools}
\usepackage{placeins}
\usepackage{xcolor}
\usepackage{soul}
\usepackage[outline]{contour}
\usepackage{amssymb}
\usepackage{wasysym}
\usepackage{geometry}
\usepackage{hvfloat}
\usepackage{titlesec}
\usepackage{amssymb}
\usepackage{pifont}
\usepackage{mathrsfs}
\usepackage{subcaption}

\definecolor{tab:blue}{RGB}{31, 119, 180}
\definecolor{tab:red}{RGB}{200,0,0}
\definecolor{tab:magenta}{cmyk}{0,0,1,0}
\definecolor{tab:green}{RGB}{0,180,0}

\newcolumntype{P}[1]{>{\centering\arraybackslash}p{#1}}

\definecolor{babyblue}{rgb}{0.54, 0.81, 0.94}

\begin{document}

\title{Study of neutrino mass matrices with vanishing trace and one vanishing minor}

\author[a,1]{\small Sangeeta Dey}
\author[a,2]{\small Mahadev Patgiri}

\affil[a]{\footnotesize Department of Physics, Cotton University, Guwahati, India}
\affil[1]{\textit{phy1891006\_sangeeta@cottonuniversity.ac.in}}
\affil[2]{\textit{mahadevpatgiri@cottonuniversity.ac.in}}

\maketitle

\begin{abstract}

   In this work we carry out a systematic texture study of the neutrino mass matrix with the ansatzes - (i) one vanishing minor and (ii) the zero sum of the mass eigenvalues with the CP phases (henceforth vanishing trace). There are six possible textures of a neutrino mass matrix with one vanishing minor. The viability of each texture is checked with $3\sigma$ values of current neutrino data by drawing scatter plots. In our analysis we are motivated to use the ratio of solar to atmospheric mass-squared differences $R_{\nu}$ for its precise measurement (and also the atmospheric mixing angle $\theta_{23}$) to constrain phenomenologically first the Dirac CP phase $\delta$ in the range of $0^{\degree}-360^{\degree}$ for a given texture with the solutions of the constraint equations. Subsequently we employ this constrained $\delta$ to determine the range of completely unknown Majorana CP Phases ($\alpha$ and $\beta$) for all the viable textures. We also check the neutrinoless double beta decay rate, $|m_{ee}|$ and the Jarlskog invariant, $J_{cp}$ for the textures. Finally the symmetry realization of all the viable textures under the flavor symmetry group $Z_5$ via seesaw mechanism is implemented along with the FN mechanism to determine mass hierarchy structure. 
\end{abstract}

\section{INTRODUCTION}
  
   The phenomenon of neutrino oscillations i.e., the change from one flavor to another has been decisively confirmed by the results of the neutrino oscillation experiments carried out for last few decades. The neutrino oscillation theory predicts massive neutrinos and flavor mixing. Many neutrino oscillation experiments \cite{adamson2014combined,abe2011indication,ayres2005nova, an2012observation, ahn2012observation} have entered the regime of precision measurement of the three mixing angles $(\theta_{12},\theta_{23},\theta_{13})$ and two mass squared differences $(\Delta m_{12}^2, \Delta m_{23}^2)$. The ordering of neutrino masses is not yet known but can be probed in the experiments viz., JUNO experiment\cite{an2016neutrino}, long baseline experiment with Hyper-Kamiokande detector and J-PARK accelerator\cite{hyper2015physics}, DUNE experiment\cite{abi2020long}. The absolute scale of neutrinos is also not yet experimentally known but the information about the upper bounds obtained from KATRIN\cite{wolf2010katrin}, GERDA\cite{abt2004new}, EXO\cite{danilov2000detection}, KamLAND-ZEN \cite{gando2016search} experiments and some cosmological observations indicate the sub-eV scale.\par
   Theoretical understanding of origin of such small neutrino masses and large mixings is a very important issue to be addressed in leptonic sector of particle physics. In three neutrino flavour scheme, the neutrinos are described by a symmetric $3\times3$ mass matrix $M_{\nu}$ which is diagonalized by the PMNS mixing matrix, $V_{PMNS}$ giving the mass eigenvalues of light neutrinos. The textures of the effective neutrino mass matrix have been investigated with different proposals in the literature, e.g.,the vanishing minors\cite{lavoura2005zeros,lashin2008zero,lashin2009one,dev2010neutrino,verma2012non,dev2011near,liao2014dual}, zero elements \cite{fritzsch2011two,barranco2012neutrino,liao2013seesaw,liao2013one,lavoura2013reproducing, gautam2015neutrino,liao2015neutrino,kitabayashi2020texture,gautam2018trimaximal,frampton2002zeroes,xing2002texture}, equality of elements/minors \cite{dev2010phenomenological,dev2013neutrino,dev2010parallel,kaneko2005hybrid}, zero trace\cite{he2003neutrino,singh2018texture}, zero determinant\cite{branco2003removing} etc., that are phenomenologically viable on the light of the current neutrino data. Such texture study restricts the possible structures of neutrino mass matrix and also reduces the free parameters. From the point of model building, this approach is useful and economical. Again, the canonical seesaw mechanism is a simple and theoretically appealing framework beyond the Standard Model of particle physics to generate tiny masses and large mixing of observable neutrinos. In this framework, $M_{\nu}$ is built from two more fundamental mass matrices: (i) the Dirac neutrino mass matrix $M_{D}$ and (ii) the heavy right-handed Majorana neutrino mass matrix $M_{R}$. We adopt a principle that any texture of $M_{\nu}$ acquired is a result of the combined effect of the textures of $M_{D}$ and $M_{R}$ via canonical seesaw formula. The lepton sector is not yet completely known. In the seesaw framework, neutrinos would be completely described by three masses $(m_{1}, m_{2},m_{3})$, three mixing angles $(\theta_{12},\theta_{13},\theta_{23})$ and one Dirac CP phase $\delta$ and two Majorana CP phases $(\alpha,\beta)$. Currently we have the experimental data on two mass squared differences and hence the ratio of these two mass squared differences $R_{\nu}$, three mixing angles, the strength of CP violation, the Jarlskog invariant $J_{CP}$ and the neutrinoless double beta decay rate $|m_{ee}|$. Now in phenomenology, the texture study of $M_{\nu}$ maybe a useful tool for predicting values of the other unknown parameters on the basis of currently available data. Such conditions on $M_{\nu}$ theoretically indicate some underlying flavor symmetry and hence the origin of such textures becomes important in model building.\par 
  In the present work we intend to explore the texture of neutrino mass matrices with two ansatzes imposing simultaneously: (i) one vanishing minor and (ii) zero sum of the mass eigenvalues with the CP phases \cite{rodejohann2004neutrino}, henceforth it will be termed as the vanishing trace. The following are the primary motivations of considering these two ansatzes in our work :\par 
  (a) In seesaw mechanism neutrino mass matrix is given by $M_\nu= -M_D M_R^{-1}M_D^T$ obtained from the Dirac mass matrix $M_D$ and the heavy right-handed Majorana mass matrix $M_R$ which are considered to be  more fundamental. Now the zeros in $M_D$ and $M_R$ propagate to $M_\nu$ and manifests as its vanishing minor(s) or texture zero(s) via seesaw formula. On the otherhand, the zeros in $M_D$ and $M_R$ represent the underlying flavor symmetry that may be realized by the discrete symmetry group $Z_N$ \cite {grimus2005neutrino, grimus2004symmetry}. These $Z_N$ is a subgroup of $U(1)$ Abelian gauge group, which gives a strong theoretical foundation in this approach.\par 
  (b) Some of the non-oscillation experiments viz., neutrinoless double beta decay, tritium beta decay end point spectrum etc. can measure the absolute mass scale directly, whereas the oscillation experiments can measure the mass squared differences of neutrinos known as solar and atmospheric mass splittings, and ordering of neutrino masses. It is also noted that authors in their paper \cite{black2000complementary} showed that traceless condition enabled one to calculate the absolute masses of neutrinos in normal hierarchy (NH) or in inverted hierarchy (IH) mass pattern from the current neutrino data. 
  
  We have six possible textures of $M_{\nu}$ each having one vanishing minor and vanishing trace. Each texture has two simultaneous constraint equations in two variables defined from the ratio of mass eigenvalues with CP phases. With the solutions we plot $R_{\nu}$ (and also $\theta_{23}$) versus $\delta$ to check the viability of the particular texture. The $3\sigma$ values of $R_{\nu}$ and $\theta_{23}$ either restrict a texture to the sub-range of $\delta$ out of the full range $0^{\degree}-360^{\degree}$ or completely rule out. Then we explore the range of Majorana CP phases by plotting $\alpha$ and $\beta$ against the allowed range of $\delta$ for viable cases. The viability of those allowed textures are further checked in the light of $|m_{ee}|$ and $J_{cp}$. The successful textures in our proposed investigations are subject to the symmetry realization under the flavor group $Z_5$ in seesaw mechanism which is further augmented by the Froggatt-Nielsen (FN) mechanism \cite{froggatt1979hierarchy,han2017neutrino} to know the mass pattern. \par
   
   The paper is organized as follows: In Sec.II, we present the framework of the neutrino mass matrix having one vanishing minor and vanishing trace, followed by the texture analysis in Sec.III. In Sec.IV we have done the symmetry realization of viable textures. Finally we present results and discussion in Sec.V.
\section{NEUTRINO MASS MATRIX FRAMEWORK} 
At first we consider the neutrino mass matrix $M_{\nu}$ as
 \begin{equation}\label{(1)}
 M_{\nu} =               
 V\begin{pmatrix}        
 m_1 & 0 & 0 \\           
 0 & m_2 & 0 \\           
 0 & 0 & m_3 \\           
 \end{pmatrix}V^T,    
\end{equation}          
where ($m_1$, $m_2$, $m_3$) are the neutrino mass eigenvalues and V is the diagonalising PMNS matrix with the following parametrization\cite{wang2013neutrino} in the basis of the diagonal charged lepton mass matrix:  
  \begin{center}
  $V = UP_{\nu}$     
  \end{center}  
  \begin{equation}\label{(2)}
   =\begin{pmatrix}        
 c_{12}c_{13} & c_{13}s_{12} & s_{13}e^{-i\delta}\\           
 -s_{12}c_{23}-c_{12}s_{13}s_{23}e^{i\delta} & c_{12}c_{23}-s_{12}s_{13}s_{23}e^{i\delta} & c_{13}s_{23} \\           
 s_{23}s_{12}-c_{12}c_{23}s_{13}e^{i\delta} & -c_{12}s_{23}-c_{23}s_{12}s_{13}e^{i\delta} & c_{13}c_{23} \\           
 \end{pmatrix} diag(1,e^{i\alpha},e^{i(\beta+\delta)}),
 \end{equation}
 where ($\theta_{12}$, $\theta_{23}$, $\theta_{13}$) are the solar, atmospheric and reactor mixing angles respectively, $\delta$ is the Dirac CP phase and $\alpha$, $\beta$ are the Majorana CP phases. 
 
 Now we can re-cast the neutrino mass matrix in the following strategic form:
 \begin{equation}\label{(3)}
 M_{\nu} =               
 U\begin{pmatrix}        
 \lambda_1 & 0 & 0 \\           
 0 & \lambda_2 & 0 \\           
 0 & 0 & \lambda_3 \\           
 \end{pmatrix}U^T, 
 \end{equation}
where $\lambda_1=m_1$, $\lambda_2=m_2e^{2i\alpha}$, $\lambda_3=m_3e^{2i(\beta+\delta)}$.
Now using Eq.\ref{(3)} any element of the neutrino mass matrix $M_\nu$ can be expressed as
\begin{equation}\label{(4)}
(M_\nu)_{mn}=\sum_{i=1}^3 U_{mi}U_{ni}\lambda_i.
\end{equation}
The co-factors of the off-diagonal elements of the symmetric matrix $M_{\nu}$ can be written in the following form:
\begin{equation}\label{(5)}
C_{mn}=(-1)^{m+n}((M_{\nu})_{(m+1,n-1)}(M_{\nu})_{(m+2,n+1)}-(M_{\nu})_{(m+1,n+1)}(M_{\nu})_{(m+2,n+2)}),
\end{equation}
and that for diagonal elements:
\begin{equation}\label{(6)}
C_{mm}=(-1)^{2m}((M_{\nu})_{(m+1,m+1)}(M_{\nu})_{(m+2,m+2)}-(M_{\nu})_{(m+1,m+2)}(M_{\nu})_{(m+2,m+1)}).
\end{equation}
For $m+l$, $n+l > 3$, we have to take the values $(m+l)-3$, $(n+l)-3$. Here m, n can take values (1, 2, 3) and $l=1, 2$. Now we impose the condition for vanishing minor, i.e.,
\begin{equation}\label{(7)}
C_{mn}=0,\;\; C_{mm}=0.
\end{equation}
Solving Eq.\ref{(7)} we get 
\begin{equation}\label{(8)}
m_1m_2e^{2i\alpha}A_3 + m_2m_3e^{2i(\alpha+\beta+\delta)}A_1+m_3m_1e^{2i(\beta+\delta)}A_2 =0,
\end{equation}
where
\begin{equation}\label{(9)}
 A_i=(U_{pj}U_{qj}U_{rk}U_{sk}-U_{tj}U_{uj}U_{vk}U_{wk})+(j\longleftrightarrow k),
\end{equation}
here (i, j, k) is a cyclic permutation of (1, 2, 3). Therefore the two constraint equations are
\begin{equation}\label{(10)}
\lambda_1\lambda_2A_3 + \lambda_2\lambda_3A_1+\lambda_3\lambda_1A_2 =0,
\end{equation}
\begin{equation}\label{(11)}
\lambda_1+\lambda_2+\lambda_3=0.
\end{equation}
Considering $\lambda_1>0$ and defining $X=\frac{\lambda_2}{\lambda_1}$ and $Y=\frac{\lambda_3}{\lambda_1}$, the Eqs.\ref{(10)} and \ref{(11)} become
\begin{equation}\label{(1001)}
X A_3 + X Y A_1+YA_2 =0,
\end{equation}
\begin{equation}\label{(1101)}
1+X+Y=0.
\end{equation}
Solving Eq.\ref{(1001)} and Eq.\ref{(1101)} we get the following ratios 
\begin{equation}\label{(12)}
X_{\pm}=\frac{(A_3-A_1-A_2)\pm \sqrt{(A_3-A_1-A_2)^2-4A_1A_2}}{2A_1}, 
\end{equation}
\begin{equation}\label{(13)}
Y_{\pm}=\frac{(A_2-A_1-A_3)\pm \sqrt{(A_3-A_1-A_2)^2-4A_1A_2}}{2A_1}.
\end{equation}
For the solution pairs $(X_{+}, Y_{-})$ and $(X_{-}, Y_{+})$ we get the Majorana phases as 
\begin{equation}\label{(14)}
\alpha=\frac{1}{2}Arg[\frac{(A_3-A_1-A_2)\pm \sqrt{(A_3-A_1-A_2)^2-4A_1A_2}}{2A_1}],
\end{equation}
\begin{equation}\label{(15)}
\beta=\frac{1}{2}Arg[\frac{(A_2-A_1-A_3)\pm \sqrt{(A_3-A_1-A_2)^2-4A_1A_2}}{2A_1}]e^{-2i\delta}.
\end{equation}
The other two solution pairs $(X_{+}, Y_{+})$ and $(X_{-}, Y_{-})$ satisfy the constraint equations under the condition $(A_3-A_1-A_2)^2-4A_1A_2=0$. For these two solution pairs we have the Majorana phases as
\begin{equation}\label{(16)}
\alpha=\frac{1}{2}Arg[\frac{(A_3-A_1-A_2)}{2A_1}];\quad \beta=\frac{1}{2}Arg[\frac{(A_2-A_1-A_3)}{2A_1}]e^{-2i\delta}.
\end{equation}
 The ratios of the neutrino masses
 \begin{equation}\label{(17)}
 \rho=|\frac{m_2}{m_1}e^{2i\alpha}| 
 \end{equation}
 and
 \begin{equation}\label{(18)}
 \sigma=|\frac{m_3}{m_1}e^{2i\beta}|
 \end{equation}
are related to the ratio of solar and atmospheric mass-squared differences 
\begin{equation}\label{(19)}
R_\nu=\frac{\delta m^2}{\Delta m^2}=\frac{2(\rho^2-1)}{2\sigma^2-\rho^2-1},
\end{equation}
where $\delta m^2=m_2^2-m_1^2$ called the solar mass splitting and $\Delta m^2=|m_{3}^2-\frac{1}{2}(m_2^2+m_1^2)|$ the atmospheric mass splitting. For NH, the ratio $R_\nu=\frac{2\epsilon_1}{\sigma^2}$, if we consider $\rho=1+\epsilon $ for $m_1$ and $m_2$ being very close to each other with the values $0.013<\epsilon<0.017$ on $3\sigma$.  For IH, $R_\nu=\frac{2(\rho^2-1)}{\rho^2+1}$.
The $3\sigma$ values of $\Delta m_{21}^2$ and $|\Delta m_{3l}^2|$ we obtain the range $R_{\nu}=(0.026-0.035)$.\par 
The measure of CP violation i.e the Jarlskog invariant \cite{jarlskog1985commutator} is given by \par 
\begin{equation}\label{(20)}
 J_{CP}= \frac{1}{8}\sin{2\theta_{12}}\sin{2\theta_{23}}\sin{2\theta_{13}}\cos{\theta_{13}}\sin{\delta}.
\end{equation} 
The nature of the neutrino is still unknown whether it is the Dirac or the Majorana type. The observation of neutrinoless double beta decay would indicate the process of the lepton number violation and confirming the Majorana nature of neutrinos. The rate of neutrinoless double beta decay depends on the effective Majorana mass of electron neutrino: 
\begin{equation}\label{(21)}
 |m_{ee}|= |m_{1}c_{12}^2c_{13}^2+m_2s_{12}^2c_{13}^2e^{2i\alpha}+m_3s_{13}^2e^{2i\beta}|.
\end{equation} 
Various ongoing and upcoming neutrinoless double beta decay experiments such as CUORICINO \cite{arnaboldi2004first}, CUORE \cite{gironi2010performance}, GERDA \cite{abt2004new}, MAJORANA \cite{aalseth2005proposed}, SuperNEMO \cite{barabash2006double}, EXO \cite{danilov2000detection}, GENIUS \cite{klapdor2002reply}, target to achieve a sensitivity upto 0.01eV for $|m_{ee}|$. The most constraint upper limit has been set to $|m_{ee}|< 0.061-0.165$ eV at $90\%$ CL by the KamLAND-ZEN Collaboration \cite{gando2016search}. \par

\begin{table}[H]

\caption{Current neutrino oscillation parameters from global fits \cite{gonzalez2021nufit}. Here $\Delta m_{3l}^2=\Delta m_{31}^2>0$ for normal hierarchy and $\Delta m_{3l}^2=\Delta m_{32}^2<0$ for inverted hierarchy.}
\centering
\begin{tabular}{|c|c c|c c|}
\hline
Parameter & \multicolumn{2}{c|}{Normal Ordering}&\multicolumn{2}{c|}{Inverted Ordering}\\
          & best fit $\pm 1\sigma$ & 3$\sigma$ range & best fit$\pm 1\sigma$ & 3$\sigma$ range\\ 
\hline
$\theta_{12}^{\degree}$ & $33.45^{+0.77}_{-0.75}$ & $31.27-35.86$ & $33.45^{+0.78}_{-0.75}$ & $31.27-35.87$\\

\hline
$\theta_{23}^{\degree}$ & $42.1^{+1.1}_{-0.9}$ & $39.7-50.9$ & $49.0^{+0.9}_{-1.3}$ & $39.8-51.6$ \\

\hline
$\theta_{13}^{\degree}$ & $8.62^{+0.12}_{-0.12}$ & $8.25-8.98$ & $8.61^{+0.14}_{-0.12}$ & $8.24-9.02$ \\
\hline
$\delta_{cp}^{\degree}$ & $230^{+36}_{-25}$ & $144-350$ & $278^{+22}_{-30}$ & $194-345$ \\

\hline
$\Delta m_{21}^2/10^{-5}eV^2$ & $7.42^{+0.21}_{-0.20}$ & $6.82-8.04$ & $7.42^{+0.21}_{-0.20}$ & $6.82-8.04$ \\
\hline
$|\Delta m_{3l}^2|/10^{-3}eV^2$ & $2.510^{+0.027}_{-0.027}$ & $2.430-2.593$ & $2.490^{+0.026}_{-0.028}$ & $-2.574- -2.410$ \\
\hline
\end{tabular}
\end{table}

\section{ TEXTURE ANALYSIS}
We undertake the following strategy for systematic and comprehensive study of the six possible textures of vanishing minors with the collateral condition of vanishing trace.
\begin{itemize}
\item[(i)] For a given texture of $C_{ij}=0$ with vanishing trace $\sum{\lambda_{i}}=0$, we construct the Eqs.\ref{(1001)} and \ref{(1101)} with the ratios $\frac{\lambda_{2}}{\lambda_{1}}=X$ and $\frac{\lambda_{3}}{\lambda_{1}}=Y$. Since the Eq.\ref{(1001)} is having cross term of $X$ and $Y$, so we have two solutions for each of $X$ and $Y$.
Now there are four options of solution pairs viz., ($X_{+}$, $Y_{+}$), ($X_{+}$, $Y_{-}$), ($X_{-}$, $Y_{+}$) and ($X_{-}$, $Y_{-}$) for texture study to be carried out.

\item[(ii)] Using above solution pairs in Eq.\ref{(19)} via Eqs.\ref{(17)} and \ref{(18)}, we generate the random numbers for $R_{\nu}$ allowing three mixing angles $\theta_{12}$, $\theta_{13}$, $\theta_{23}$ to pick up random numbers in their corresponding $3\sigma$ values and the Dirac phase $\delta$ in the range $0^{\degree}$-$360^{\degree}$. Then we plot $R_{\nu}$ versus $\delta$ for normal and inverted mass ordering. If $R_{\nu}$ retains its values within the experimental limits, the texture is considered for further phenomenological study within this allowed range of $\delta$, otherwise rejected. The range of $\delta$ so obtained is further checked by plotting the atmospheric mixing angle $\theta_{23}$ against $\delta$ also for subsequent use.

\item[(iii)] With the phenomenologically allowed range of $\delta$ obtained via (ii), scatter plots are drawn to find the values of the Majorana phases $\alpha$ and $\beta$ which may be measured by the experiments in future.

\item[(iv)] The viable textures are further explored for the effective Majorana mass of electron neutrino $|m_{ee}|$ indicating the rate of neutrinoless double beta decay and the Jarlskog invariant, $J_{cp}$ representing the strength of CP violation in neutrino oscillations.
\end{itemize}
 To avoid making the paper loaded with a number of plots, now we choose to present the detailed analysis of two textures $C_{11}=0$ and $C_{12}=0$ only as representative cases and the results for other textures shall be summarized in the Table (II) and (III).
\subsection{\underline{Case $C_{11}=0$}}
 For this texture we have
 \begin{equation}\label{(22)}
 A_1=c_{12}^2 c_{13}^2,
 \end{equation}
 \begin{equation}\label{(23)}
 A_2=s_{12}^2c_{13}^2,
 \end{equation}\label{(24)}
 \begin{equation}
 A_3=s_{13}^2e^{2i\delta}.
 \end{equation}

Now we first consider the solution pair $(X_{+}, Y_{-})$ to plot $R_{\nu}$ (Eq.\ref{(19)}) for both NH and IH. 
\begin{figure}[H]
\begin{subfigure}[b]{0.4\textwidth}
\includegraphics[width=\textwidth]{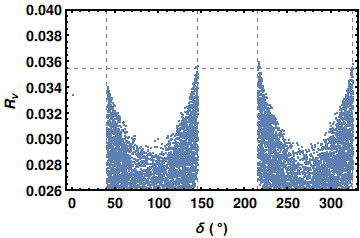}
\caption{}
\end{subfigure}
\hfill
\begin{subfigure}[b]{0.4\textwidth}
\includegraphics[width=\textwidth]{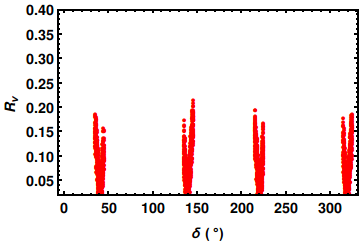}  
\caption{}
 \end{subfigure}
\caption{$R_{\nu}$ plots (a) for NH and (b) for IH.}
\label{FIG.1}
\end{figure}
In Fig.\ref{FIG.1}(a), the ratio $R_{\nu}$ lies within the allowed experimental values that constrain the Dirac phase $\delta$ in the range of $(50^{\degree},150^{\degree})\oplus(220^{\degree}, 320^{\degree})$ for NH, while in Fig.\ref{FIG.1}(b) the ratio $R_{\nu}$ lies outside the allowed range and hence the same texture is phenomenologically ruled out at 3$\sigma$ level for IH. Then with the allowed range of $\delta$ for NH, the scatter plots are drawn for $\alpha$ and $\beta$ in Fig.\ref{FIG.(2)}. From the plots we obtain the Majorana phases $\alpha$ in the range of $(-25^{\degree}, 25^{\degree})$ and $\beta$ in $(-45^{\degree}, 45^{\degree})$. \par
\begin{figure}[H]
\begin{subfigure}[b]{0.4\textwidth}
\includegraphics[width=\textwidth]{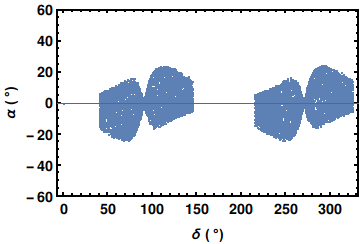}
\caption{} 
\end{subfigure}
\hfill
\begin{subfigure}[b]{0.4\textwidth}
\includegraphics[width=\textwidth]{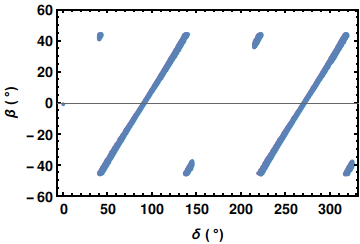} 
\caption{} 
\end{subfigure}
\caption{$\alpha$ and $\beta$  plots for NH for the pair $(X_+,Y_-)$ with $\delta=(50^{\degree},150^{\degree})\oplus(220^{\degree}, 320^{\degree})$.}
\label{FIG.(2)}
\end{figure}
 
Similar procedure is followed for the solution pairs $(X_{-},Y_{+})$,$(X_{+},Y_{+})$ and $(X_{-},Y_{-})$ of the texture but plots show that $R_{\nu}$ acquires values far beyond the experimental range. Hence all these solutions of the texture are ruled out.
\par
\begin{figure}[H]
\centering
\begin{subfigure}[b]{0.4\textwidth}
\includegraphics[width=\textwidth]{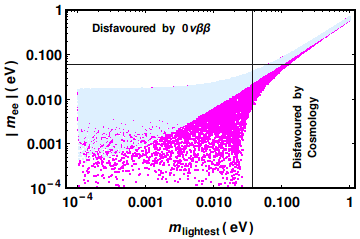}  
\caption{}
\end{subfigure}
\hfill
\begin{subfigure}[b]{0.4\textwidth}
\includegraphics[width=\textwidth]{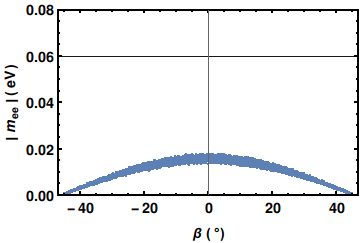}  
\caption{}
\end{subfigure}
\caption{$|m_{ee}|$ plots for $(X_{+}, Y_{-})$ for NH versus $m_{lightest}$ and $\beta$.\;\; Region in \fcolorbox{magenta}{magenta}{\rule{0pt}{2.5pt}\rule{2.5pt}{0pt}}\quad indicates the experimental bounds and \fcolorbox{babyblue}{babyblue}{\rule{0pt}{2.5pt}\rule{2.5pt}{0pt}}\quad shows the allowed region for $(X_{+}, Y_{-})$ for NH.}
 \label{FIG.3}
\end{figure}
\begin{figure}[H]
\centering
\begin{subfigure}[b]{0.4\textwidth}
\includegraphics[width=\textwidth]{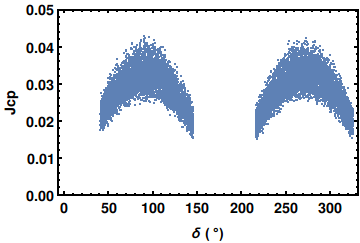}
\end{subfigure}
\caption{$J_{cp}$ plot for NH for the solution pair $(X_{+}, Y_{-})$.}
\label{FIG.4}
\end{figure}

Now to explore further phenomenolgy of the texture, $|m_{ee}|$ - $m_{lightest}$ and $|m_{ee}|$-$\beta$ are plotted for neutrinoless double beta decay where the mass of the lightest neutrino is bound within $0.037$ eV and $0.042$ eV for NH and IH respectively at 95$\%$ confidence \cite{stocker2021strengthening}. We also plot $J_{cp}$-$\delta$ for CP violation. From Fig.\ref{FIG.3} we observe that $|m_{ee}|$ lies within the experimental bounds which are similar results as in the paper \cite{chen2018realistic}. Again, in Fig.\ref{FIG.4}, we find $J_{cp}$ within the range $(0.018-0.04)$. Thus the $C_{11}=0$ is found viable under the phenomenological study for normal mass ordering in case of the solution pair $(X_{+}, Y_{-})$ of the texture.

\subsection{\underline{Case $C_{12}=0$}}
For this texture $A_1$, $A_2$ and $A_3$ are the following:
\begin{equation}\label{(25)}
A_{1}=c_{12}s_{12}c_{23}c_{13}+c_{12}^2c_{13}s_{13}s_{23}e^{-i \delta},
\end{equation}
\begin{equation}\label{(26)}
A_{2}=-c_{12}s_{12}c_{23}c_{13}+s_{12}^2c_{13}s_{13}s_{23}e^{-i\delta},
\end{equation}
\begin{equation}\label{(27)}
A_{3}=-s_{23}s_{13}c_{13}e^{i\delta}.
\end{equation}

Now we consider the solution pair $(X_{+},Y_{-})$ for the texture. 
\begin{figure}[H]
\centering
\begin{subfigure}[b]{0.4\textwidth}
\includegraphics[width=\textwidth]{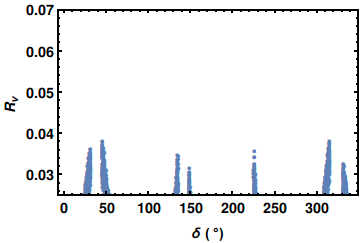}
\end{subfigure}
\caption{$R_{\nu}$ plot for NH for the solution pair $(X_{+}, Y_{-})$.}
\label{FIG.5}
\end{figure}
Fig.\ref{FIG.5} shows that $\delta$ is constrained in the range $(20^{\degree},31^{\degree})\oplus(45^{\degree}, 55^{\degree})\oplus (128^{\degree}, 135^{\degree})\oplus(148^{\degree}, 152^{\degree})\oplus(225^{\degree},231^{\degree})\oplus(300^{\degree},315^{\degree})\oplus(331^{\degree},342^{\degree})$ for NH. It is found that for IH the ratio $R_{\nu}$ lies outside the allowed experimental range. 
\par
\begin{figure}[H]
\begin{subfigure}[b]{0.4\textwidth}
\includegraphics[width=\textwidth]{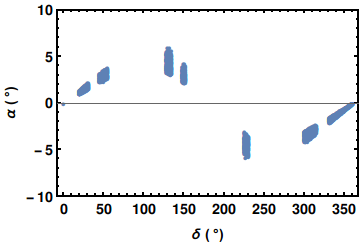}
\caption{}
\end{subfigure}
\hfill
\begin{subfigure}[b]{0.4\textwidth}
\includegraphics[width=\textwidth]{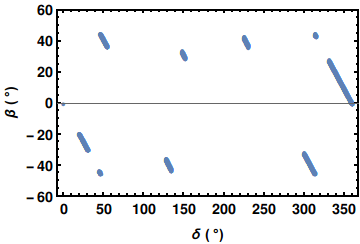} 
\caption{} 
\end{subfigure}
\caption{$\alpha$ and $\beta$ plots for NH for the solution pair $(X_{+}, Y_{-})$where $\delta$ lies within the range $(20^{\degree},31^{\degree})\oplus(45^{\degree}, 55^{\degree})\oplus (128^{\degree}, 135^{\degree})\oplus(148^{\degree}, 152^{\degree})\oplus(225^{\degree},231^{\degree})\oplus(300^{\degree},315^{\degree})\oplus(331^{\degree},342^{\degree})$.}
\label{FIG.6}
\end{figure}

On plotting the graphs for $\alpha$ and $\beta$ in Fig.\ref{FIG.6}, we obtain $\alpha=(-6^{\degree}, 6^{\degree})$  and $\beta=(-45^{\degree}, -20^{\degree})\oplus(0, 45^{\degree})$.
\par
\par
Now we consider the solution pair $(X_{-}, Y_{+})$. 
\begin{figure}[H]
\centering
\begin{subfigure}[b]{0.4\textwidth}
\includegraphics[width=\textwidth]{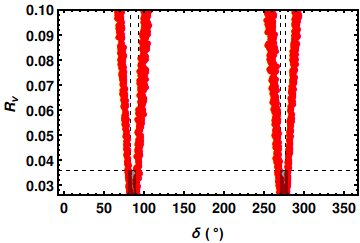} 
\end{subfigure}
\caption{$R_{\nu}$ plot for IH with the solution pair $(X_{-}, Y_{+})$}
\label{FIG.7}
\end{figure}
 The plot for $R_{\nu}$ in Fig.\ref{FIG.7} shows that it lies within the experimental range for $\delta=(82^{\degree},92^{\degree})\oplus(270^{\degree}, 276^{\degree})$ for IH only, while the texture is ruled out for NH. In Fig.\ref{FIG.8} we find highly constrained values of $\alpha$ and $\beta$ as $(-6^{\degree}, -3^{\degree})\oplus(3^{\degree},6^{\degree})$ and $(-45^{\degree}, -35^{\degree})\oplus(35^{\degree},45^{\degree})$ respectively for IH.
\begin{figure}[H]
\begin{subfigure}[b]{0.4\textwidth}
\includegraphics[width=\textwidth]{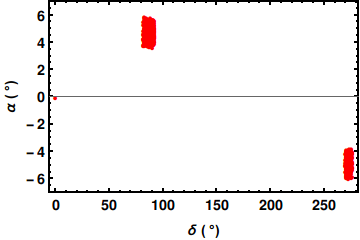}
\caption{}
\end{subfigure}
\hfill
\begin{subfigure}[b]{0.4\textwidth}
\includegraphics[width=\textwidth]{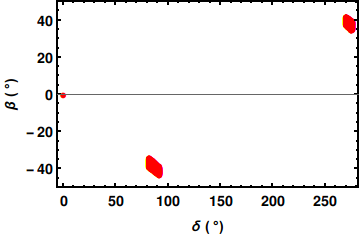}  
\caption{}
\end{subfigure}
\caption{$\alpha$ and $\beta$ plots for IH for the solution pair $(X_{-}, Y_{+})$}
\label{FIG.8}
\end{figure}

\par 
Similar prescription has been used for the solution pairs $(X_{+}, Y_{+})$ and $(X_{-}, Y_{-})$.
\begin{figure}[H]
\begin{subfigure}[b]{0.4\textwidth}
\includegraphics[width=\textwidth]{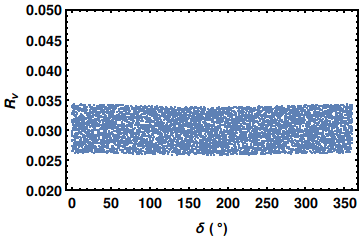}
\caption{}
\end{subfigure}
\hfill
\begin{subfigure}[b]{0.4\textwidth}
\includegraphics[width=\textwidth]{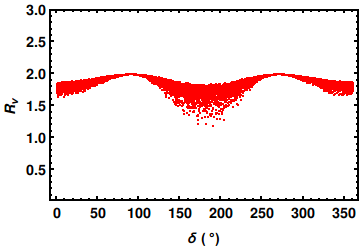} 
\caption{} 
\end{subfigure}
\caption{$R_{\nu}$ plots for the pair $(X_{+}, Y_{+})$ and $(X_{-}, Y_{-})$. The left plot is for NH and the right plot for IH.}
\label{FIG.9}
\end{figure}
Fig.\ref{FIG.9} shows that the pairs $(X_+, Y_+)$ and $(X_-, Y_-)$ are viable only for NH in the entire range of $\delta$ i.e, $(0,360^{\degree})$. Fig.\ref{FIG.10} gives the Majorana phases $\alpha=(-10^{\degree}, 10^{\degree})$ and $\beta=(-50^{\degree},50^{\degree})$.
\begin{figure}[H]
\begin{subfigure}[b]{0.4\textwidth}
\includegraphics[width=\textwidth]{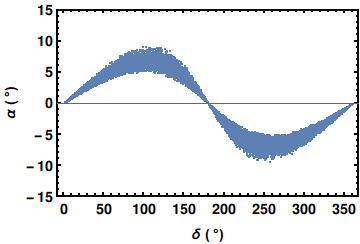}
\caption{}
\end{subfigure}
\hfill
\begin{subfigure}[b]{0.4\textwidth}
\includegraphics[width=\textwidth]{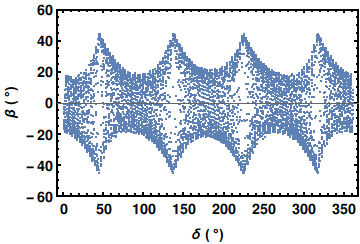} 
\caption{} 
\end{subfigure}
\caption{$\alpha$ and $\beta$ plots for the pair $(X_{+}, Y_{+})$ and $(X_{-}, Y_{-})$.}
\label{FIG.10}
\end{figure}
Now we have plotted $|m_{ee}|$ and $J_{cp}$ for viable cases only in Figs.\ref{FIG.11} and \ref{FIG.12}. In Fig.\ref{FIG.11}, we observe that $|m_{ee}|$ lies within the allowed range and Fig.\ref{FIG.12} gives $J_{cp}=(0.01, 0.04)$ for NH for $(X_{+}, Y_-)$; $(0.023, 0.04)$ for IH for $(X_{-}, Y_+)$, and $(0, 0.04)$ for NH for both $(X_{+}, Y_{+})$ and $(X_{-}, Y_{-})$. 
\begin{figure}[H]
\begin{subfigure}[b]{0.3\textwidth}
\includegraphics[width=\textwidth]{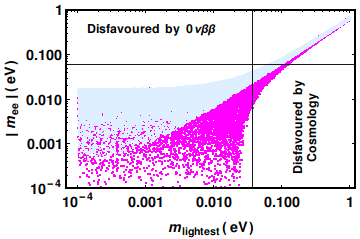}
\caption{}
\end{subfigure}
\hfill
\begin{subfigure}[b]{0.3\textwidth}
\includegraphics[width=\textwidth]{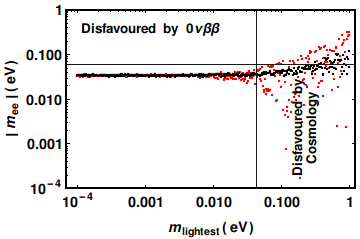}  
\caption{}
\end{subfigure}
\hfill
\begin{subfigure}[b]{0.3\textwidth}
\includegraphics[width=\textwidth]{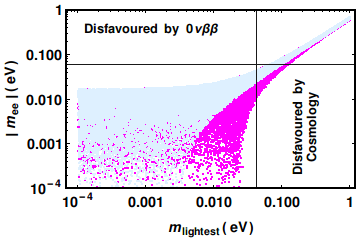}  
\caption{}
\end{subfigure}
\hfill
\begin{subfigure}[b]{0.3\textwidth}
\includegraphics[width=\textwidth]{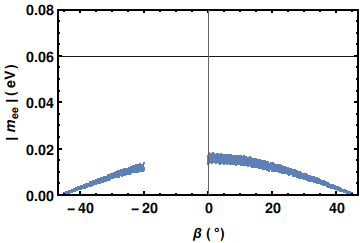}
\caption{}
\end{subfigure}
\hfill
\begin{subfigure}[b]{0.3\textwidth}
\includegraphics[width=\textwidth]{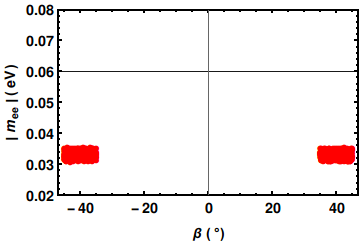}  
\caption{}
\end{subfigure}
\hfill
\begin{subfigure}[b]{0.3\textwidth}
\includegraphics[width=\textwidth]{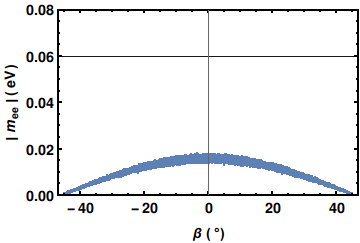}  
\caption{}
\end{subfigure}
\caption{$|m_{ee}|$ plots for $(X_{+}, Y_{-})$, $(X_{-}, Y_{+})$, $(X_{+}, Y_{+})/(X_{-}, Y_{-})$.\;\; \fcolorbox{magenta}{magenta}{\rule{0pt}{2.5pt}\rule{2.5pt}{0pt}} in (a),(c)\quad and \fcolorbox{red}{red}{\rule{0pt}{2.5pt}\rule{2.5pt}{0pt}}\quad in (b) shows the experimental bounds and \fcolorbox{babyblue}{babyblue}{\rule{0pt}{2.5pt}\rule{2.5pt}{0pt}}\quad \fcolorbox{black}{black}{\rule{0pt}{2.5pt}\rule{2.5pt}{0pt}}\quad indicates the allowed range of $|m_{ee}|$ for NH and IH for the solution pairs $(X_{+}, Y_{-})$, $(X_{-}, Y_{+})$, $(X_{+}, Y_{+})/(X_{-}, Y_{-})$.}
\label{FIG.11}
\end{figure}

\begin{figure}[H]
\begin{subfigure}[b]{0.3\textwidth}
\includegraphics[width=\textwidth]{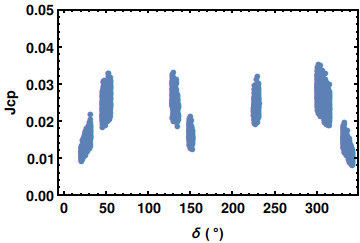}
\caption{}
\end{subfigure}
\hfill
\begin{subfigure}[b]{0.3\textwidth}
\includegraphics[width=\textwidth]{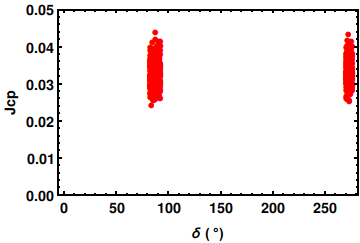}  
\caption{}
\end{subfigure}
\hfill
\begin{subfigure}[b]{0.3\textwidth}
\includegraphics[width=\textwidth]{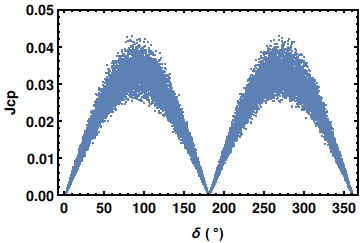} 
\caption{} 
\end{subfigure}
\caption{$J_{cp}$ plots (a), (b) and (c) for $(X_{+}, Y_{-})$, $(X_{-}, Y_{+})$, $(X_{+}, Y_{+})/(X_{-}, Y_{-})$ respectively.}
\label{FIG.12}
\end{figure}
\begin{figure}[H]
\begin{subfigure}[b]{0.3\textwidth}
\includegraphics[width=\textwidth]{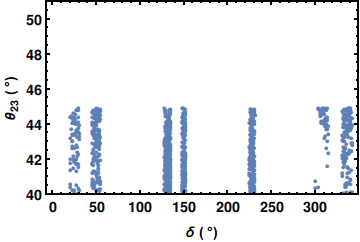}
\caption{}
\end{subfigure}
\hfill
\begin{subfigure}[b]{0.3\textwidth}
\includegraphics[width=\textwidth]{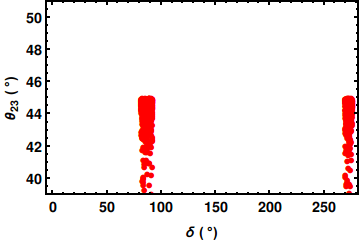}  
\caption{}
\end{subfigure}
\hfill
\begin{subfigure}[b]{0.3\textwidth}
\includegraphics[width=\textwidth]{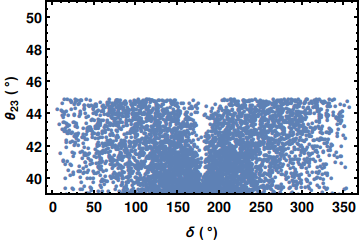} 
\caption{} 
\end{subfigure}
\caption{$\theta_{23}$ plots (a), (b) and (c) for $(X_{+}, Y_{-})$, $(X_{-}, Y_{+})$, $(X_{+}, Y_{+})/(X_{-}, Y_{-})$ respectively.}
\label{FIG.13}
\end{figure}
All the remaining textures $C_{13}=0$, $C_{22}=0$, $C_{23}=0$ and $C_{33}=0$ have been examined following our procedure of analysis. The detailed analysis are not shown in this paper, but the results are presented in the Table(II) and (III). We also checked the atmospheric mixing angle $\theta_{23}$ plotted against $\delta$ for all other viable textures and the range is always found in $(40^{\degree}$, $45^{\degree})$.
\begin{table}[H]
\small
\caption{Viable cases under normal hierarchy(NH), inverted hierarchy(IH) and neutrinoless double beta decay.}
\centering
\resizebox{11cm}{!}{\begin{tabular}{|c|c c|c c|c c|c|}
\hline
Case & \multicolumn{2}{c|}{$(X_+,Y_-)$}&\multicolumn{2}{c|}{$(X_-,Y_+)$}&\multicolumn{2}{c|}{$(X_+,Y_+)$/$(X_-,Y_-)$}& {Neutrinoless Double}\\
{} & {} & {} & {} & {} & {} & {} & Beta Decay\\ 
\hline
{} & NH & IH & NH & IH & NH & IH & {} \\
\hline
$C_{11}$ & \checkmark & x & x & x & x & x & All the viable\\ 

$C_{12}$ & \checkmark & x & x & \checkmark & \checkmark & x & cases are allowed \\

$C_{13}$ & x & x & x & x & \checkmark & x & {}\\

$C_{22}$ & \checkmark & x & \checkmark & x & x & x & {} \\

$C_{23}$ & x & x & x & x & x & x & {}\\
$C_{33}$ & \checkmark & x & x & x & x & x & {} \\

\hline
\end{tabular}}
\end{table}

\begin{table}[H]

\caption{Allowed ranges of CP phases  $\delta$, $\alpha$,\;$\beta$, $|m_{ee}|$ and $J_{cp}$ for the viable cases. }
\centering
\resizebox{\columnwidth}{!}{\begin{tabular}{|c|c c|c c|c c|}
\hline
Case & \multicolumn{2}{c|}{$(X_+, Y_-)$}&\multicolumn{2}{c|}{$(X_-, Y_+)$}&\multicolumn{2}{c|}{$(X_+, Y_+)$/$(X_-, Y_-)$}\\
\hline
{} & NH & IH & NH & IH & NH & IH\\
\hline
{}       & $\delta=(50^{\degree},150^{\degree})\oplus(220^{\degree}, 320^{\degree})$ & - & - & - & - & -\\
$C_{11}$ & $\alpha=(-25^{\degree}, 25^{\degree})$ & - & - & - & - & -\\                                      
{}       & $\beta=(-45^{\degree}, 45^{\degree})$  & - & - & - & - & -\\
{}       & $|m_{ee}|=(0, 0.02)$ eV   & - & - & - & - & -\\
{}       & $J_{cp}=(0.018, 0.04)$ & - & - & - & - & -\\
\hline
{}       & $\delta=(20^{\degree},31^{\degree})\oplus(45^{\degree}, 55^{\degree})$ & - & - & $\delta=(82^{\degree},92^{\degree})$ & $\delta=(0, 360^{\degree})$ & -\\
{}       & $\oplus (128^{\degree}, 135^{\degree})\oplus(148^{\degree}, 152^{\degree})$ & {} & {} & $\oplus (270^{\degree}, 276^{\degree})$ & {} & {}\\
$C_{12}$ & $\oplus(225^{\degree},231^{\degree})\oplus(300^{\degree},315^{\degree})$ & {} & {} & {} & {} & {} \\
{}       & $\oplus(331^{\degree},342^{\degree})$ & {} & {} & {} & {} & {} \\
{} & $\alpha=(-6^{\degree}, 6^{\degree})$ & - & - & $\alpha=(3^{\degree}, 6^{\degree})$ & $\alpha=(-10^{\degree}, 10^{\degree})$  & - \\
{} & {} & - & - & $\oplus(-6^{\degree}, -3^{\degree})$ & {}  & - \\
{} & $\beta=(-45^{\degree},-20^{\degree})\oplus(0, 45^{\degree})$ & - & - & $\beta=(-45^{\degree}, -35^{\degree})$ & $\beta=(-50^{\degree}, 50^{\degree})$ &- \\
{} & {} & - & - & $\oplus(35^{\degree}, 45^{\degree})$ & {}  & - \\
{} & $|m_{ee}|=(0, 0.02)$ eV & -& - & $|m_{ee}|=(0.03, 0.035)$ eV & $|m_{ee}|=(0, 0.02)$ eV & -\\
{} & $J_{cp}=(0.01, 0.04)$ & - & - & $J_{cp}=(0.023,0.04)$ & $J_{cp}=(0, 0.04)$ & -\\
\hline
{} & {} & - & - & - & $\delta=(55^{\degree},130^{\degree})$ & -\\
{} & {} & - & - & - & $\oplus(230^{\degree},310^{\degree})$ &  -\\
$C_{13}$ & - & - & - & - & $\alpha=(-9^{\degree}, -3^{\degree})$ & -\\
{}     & - & - & - & - & $\oplus(3^{\degree}, 9^{\degree})$ & -\\ 
{} & - & - & - & - & $\beta=(-50^{\degree}, 50^{\degree})$ & -\\
{} & - & - & - & - & $|m_{ee}|=(0, 0.02)$ eV & - \\
{} & - & - & - & - & $J_{cp}=(0, 0.04)$ & -\\
\hline
{}       & $\delta=(40^{\degree}, 90^{\degree})\oplus(230^{\degree}, 279^{\degree})$ & - & $\delta=(0, 30^{\degree})\oplus(196^{\degree}, 210^{\degree})$ & - & - & -\\ 
{}       & $\oplus(310^{\degree}, 350^{\degree})$  & {} & $\oplus(290^{\degree}, 335^{\degree})$ & - & - & -\\
$C_{22}$ & $\alpha=(-45^{\degree},-35^{\degree})\oplus(0, 45^{\degree})$ & - & $\alpha=(-45^{\degree}, 45^{\degree})$ & - & - & -\\
{}       & $\beta=(-45^{\degree}, 45^{\degree})$  & - & $\beta=(-50^{\degree}, 50^{\degree})$  & -  & - & - \\
{}       & $|m_{ee}|=(0, 0.02)$ eV  & - & $|m_{ee}|=(0, 0.02)$ eV & - & - & -\\
{} & $J_{cp}=(0, 0.04)$ & - & $J_{cp}=(0, 0.04)$ & - & - & -\\
\hline
{}       & $\delta=(90^{\degree}, 265^{\degree})$ & - & - & - & - & -\\
$C_{33}$ &  $\alpha=\beta=(-45^{\degree}, 45^{\degree})$ &  - &  - & - & - & -\\
{}       & $|m_{ee}|=(0, 0.02)$ eV & - & - & - & - & - \\ 
{} & $J_{cp}=(0, 0.04)$ & - & - & - & - & -\\
\hline
\end{tabular}}
\newline\newline
\end{table}

\section{SYMMETRY REALIZATION}
 
The most appealing theoretical approach for generating tiny masses of the light left-handed neutrinos is the seesaw mechanism with the following formula (Type-I):
\begin{equation}
M_{\nu}=-M_{D}M_{R}^{-1}M_{D}^{T}.
\end{equation} 
We carry out a systematic study to realize all the viable textures of $M_{\nu}$ in our present work, by means of the type-I seesaw mechanism with an Abelian flavor symmetry. Again, zero textures or vanishing minors of $M_{\nu}$ fundamentally origin from the zero textures $M_{D}$ and $M_{R}$ through seesaw mechanism. It is possible to enforce zero in an arbitrary entry of a fermion mass matrix by means of an Abelian flavor symmetry\cite{grimus2004symmetry}. We also note that in the lepton sector of the Standard Model, there are three right-handed charged-lepton singlets, $l_{Ri}$ and three left-handed lepton doublets, $D_{Li}$, ($i=1, 2, 3$). Further, for seesaw mechanism, three right-handed neutrino singlets $\nu_{Ri}$ are required to add. Now to build fermion mass matrices $M_{l}$, $M_{D}$ and $M_{R}$ under an Abelian flavor symmetry group, for each non-zero entry of $M_{l}$ or $M_{D}$ one needs one Higgs doublet, with appropriate transformation properties under the symmetry group, connecting two fermion multiplets corresponding to that entry, and for each non-zero entry of $M_{R}$, one requires one scalar singlet with appropriate transfomation properties under the group. Conversely, without admitting a required Higgs doublet or scalar singlet, a zero in an entry of a fermion mass matrix can be enforced.\par

We present here a detailed analysis how to realize the structure of the viable neutrino mass matrices presented in the Table II with one vanishing minor. We implement $Z_5$ Abelian flavor symmetry group to enforce zeros in fermion mass matrices. A $Z_5$ consists of the group elements \par
\begin{center}
$(1,\omega,\omega^2, \omega^3, \omega^4)$
\end{center} 
where $\omega = e^{\frac{i2\pi}{5}}$ is the generator of the group.\par

Additionally the Froggatt-Nielsen (FN) mechanism\cite{froggatt1979hierarchy,xing2020flavor} is also augmented to determine the hierarchies between neutrino masses. The FN mechanism is such an appealing approach that can explain the hierarchical structures of quarks and charged leptons. The basic idea of this mechanism is to introduce $U(1)$ global symmetry and invoke an $SU(2)_L$ singlet scalar field $\Phi$ known as a flavon field that acquires the vacuum expectation value (VEV) and breaks the $U(1)$ symmetry. This symmetry breaking is communicated to the fermions so that their effective coupling matrix elements can be expanded in powers of small positive parameter $\epsilon=\frac{<\Phi>}{\Lambda}$ with $\Lambda$, the corresponding energy scale of flavor dynamics. Thus the hierarchical textures of fermion masses can  intuitively be interpreted as powers of this expansion parameters $\epsilon$. This is the most striking feature of the FN mechanism. \par 
The Lagrangian which is responsible for the generation of the lepton masses and the hierarchy of the mass matrices arising from the FN mechanism can be written as  
\begin{align}
\begin{split}
\mathcal{L}={}&(\frac{<\Phi>}{\Lambda})^{Q_{D_{Li}}+Q_{l_{Rj}}}Y_{ij}^{(k)}\overline{D}_{Li}\phi_{k}l_{Rj}+(\frac{<\Phi>}{\Lambda})^{Q_{D_{Li}}+Q_{\nu_{R j}}}Y_{ij}^{(k)}\overline{D}_{Li}\tilde{\phi}_{k}\nu_{Rj}\\&+(\frac{<\Phi>}{\Lambda})^{Q_{\nu_{Ri}}+Q_{\nu_{R_{j}}}}Y                                                                                                                                                                                                                                                                                                                                                                                                                                                                                                                                                                                                                                                                                                                                                                                                                                                                                                                                                                                                                                                                                                                                                                                                                                                                                                                                                                                                                                                                                                                                                                                                                                                                                                                                                                                                                                                                                                                                                                                                                                                                                                                                                                                                                                                                                                                                                                                                                                                                                                                                                                                                                                                                                                                                                                                                                                                                                                                                                                                                                                                                                                                                                                                                                                                                                                                                   _{ij}^{(k)}\chi_{k}\overline{\nu}_{Ri}\nu_{Rj}+h.c.
\end{split}
\end{align}
The $Q_\alpha(\alpha=D,l_R,\nu_R$) are the FN charges for the SM fermion ingredients under which different generations may be charged differently. The flavon $\Phi$ obtains the vacuum(VEV) $<\Phi>$ that breaks the FN symmetry.
For all the cases, we assign the FN charges for the Lepton sector as\par
\begin{center}
$\overline{D}_{1, 2, 3}:(a+1, a, a)$,\par
$l_{R 1,2,3}:(0,1,2)$,\par
$\nu_{R 1,2,3}:(d,c,b)$.\par
\end{center}

Here a comment is in order. The tracelessness of $M_{\nu}$ does not speak much about the texture of $M_{\nu}$ (e.g, possible zeroes) which is supposed to be result of some deeper theory \cite{black2000complementary}. On the otherhand, the vanishing minor results in due to the seesaw mirroring between $M_{\nu}$ and $M_{R}$ with diagonal $M_{D}$ \cite{xing2020flavor}. Thus one zero texture of $M_R$ with diagonal $M_{D}$ manifests as vanishing minor of the corresponding element of $M_{\nu}$ and hence the symmetry realization under $Z_{N}$ is actionable for vanishing minor only.\par

\;\;
\;\;
\;\;
\;\;
\;\;

\underline{Symmetry realization for $C_{11}=0$}\par
\;\;
\;\;
We consider the following $M_R$ and $M_D$ for vanishing minor of the (1,1) element of $M_\nu$.\par
\begin{center}
 $M_R =               
 \begin{pmatrix}        
 0 & \xi & \zeta\\           
 \xi & \eta & \upsilon \\           
 \zeta & \upsilon & \kappa \\           
 \end{pmatrix},
 \;
 M_D =               
 \begin{pmatrix}        
 x & 0 & 0\\           
 0 & y & 0 \\           
 0 & 0 & z \\           
 \end{pmatrix}$,
 \end{center}
 \begin{equation}\label{(41)}
  M_\nu =-M_DM_R^{-1}M_D^{T}=              
 \frac{1}{\Gamma}\begin{pmatrix}        
 (-\upsilon^2+\eta\kappa)x^2 & (\zeta-\xi y)xy & (-\zeta \eta+\xi\upsilon)xz\\           
 (\zeta-\xi y)xy & -\zeta^2 y^2 & \xi\zeta yz \\           
 (-\zeta \eta+\xi\upsilon)xz & \xi\zeta yz & - \xi^2 z^2 \\            
 \end{pmatrix},
 \end{equation}
 where $\Gamma=-\zeta\eta^2+2\xi\zeta\upsilon-\xi^2\kappa$.

 On implementing $Z_5$ symmetry, the fields of the relevant particles transform as:

\begin{equation}
\begin{aligned}
\nu_{R1}\rightarrow \omega^3\nu_{R1},\;\; & \overline{D}_{L1}\rightarrow \omega^2\overline{D}_{L1},& l_{R1}\rightarrow \omega^3 l_{R1}\\
\nu_{R2}\rightarrow \omega^2\nu_{R2},\;\; & \overline{D}_{L2}\rightarrow\omega^3\overline{D}_{L2}, & l_{R2}\rightarrow \omega^2 l_{R2}\\
\nu_{R3}\rightarrow \nu_{R3},\;\;\; & \overline{D}_{L3}\rightarrow \overline{D}_{L3},& l_{R3}\rightarrow l_{R3}
\end{aligned}
\end{equation}

Here $D_{Li}, l_{Rj}, \nu_{Ri},(i,j=1,2,3)$ represents the $SU(2)_L$ doublets, the RH $SU(2)_L$ singlets and the RH neutrino singlets respectively.

Forming the required bilinears dictated by $Z_5$ symmetry we obtain\\
 $\nu_{Ri}^T\nu_{Rj} =               
 \begin{pmatrix}        
 \omega & 1 & \omega^3\\           
 1 & \omega^4 & \omega^2 \\           
 \omega^3 & \omega^2 & 1 \\           
 \end{pmatrix},
 \; 
 \overline D_{Li}\nu_{Rj} =               
 \begin{pmatrix}        
 1 & \omega^4 & \omega^2\\           
 \omega & 1 & \omega^3 \\           
 \omega^3 & \omega^2 & 1 \\           
 \end{pmatrix},
 \;
 \overline D_{Li}l_{Rj} =               
 \begin{pmatrix}        
 1 & \omega^4 & \omega^2\\           
 \omega & 1 & \omega^3 \\           
 \omega^3 & \omega^2 & 1 \\           
 \end{pmatrix}.$\\
 
 We consider the transformation of the singlet scalars $\chi_{k}(k=1,2,3)$ which is responsible for the Majorana neutrino mass matrix $M_R$ and SM-like doublet scalar $\phi $ which is responsible for the Dirac neutrino mass matrix $M_D$ and the lepton mass matrix $M_l$ under $Z_5$ transformation as 
\begin{center} 
\begin{equation} 
\begin{aligned}
\chi_{1}\rightarrow \omega^2\chi_{1},\; \chi_2\rightarrow \omega^3\chi_2,\; \chi_3\rightarrow\omega\chi_3\\
\end{aligned}
\end{equation}
\end{center}
\begin{center}
$\phi\rightarrow\phi $
\end{center}

Now the lagrangian dictated by $Z_5$ is
 \begin{align}
 \begin{split}
 \mathcal{L}_M^{Z_5}={}&\epsilon^{d+c}m_{12}\nu_{R1}^Tc^{-1}\nu_{R2}+\epsilon^{d+b}Y_{\chi_{13}}^{(1)}\chi_{1}\nu_{R1}^Tc^{-1}\nu_{R3}+\epsilon^{2c}Y_{\chi_{22}}^{(3)}\chi_{3}\nu_{R2}^Tc^{-1}\nu_{R2}+\epsilon^{c+b}Y_{\chi_{23}}^{(2)}\chi_{2}\nu_{R2}^Tc^{-1}\nu_{R3}\\&+\epsilon^{2b}m_{33}\nu_{R3}^Tc^{-1}\nu_{R3}+\epsilon^{a+d+1}Y_{D_{11}}\overline{D}_{L1}\tilde{\phi}\nu_{R1}+\epsilon^{a+c}Y_{D_{22}}\overline{D}_{L2}\tilde{\phi}\nu_{R2}+\epsilon^{a+b}Y_{D_{33}}\overline{D}_{L3}\tilde{\phi}\nu_{R3}\\&+\epsilon^{a+1}Y_{l_{11}}\overline{D}_{L1}\phi l_{R1}+\epsilon^{a+1}Y_{l_{22}}\overline{D}_{L2}\phi l_{R2}+\epsilon^{a+2}Y_{l_{33}}\overline{D}_{L3}\phi l_{R3}.
 \end{split}
 \end{align}
 
Now we construct the mass matrix $M_R$, $M_D$ and $M_l$ as 
\begin{equation}
\resizebox{0.5 \textwidth}{!}
 {$M_R =               
 \begin{pmatrix}        
 0 & \epsilon^{d+c}m_{12} & y_{\chi_{13}}^{(1)}\chi_1\epsilon^{d+b}\\           
 \epsilon^{d+c}m_{12} & y_{\chi_{22}}^{(3)}\chi_3\epsilon^{2c} & y_{\chi_{23}}^{(2)}\chi_2\epsilon^{c+b} \\           
 y_{\chi_{13}}^{(1)}\chi_1\epsilon^{d+b} & y_{\chi_{23}}^{(2)}\chi_2\epsilon^{c+b} & \epsilon^{2b}m_{33} \\           
 \end{pmatrix}$},
 \end{equation}
 \begin{equation}
 \resizebox{0.9 \textwidth}{!}
  {$M_D =               
 \begin{pmatrix}        
y_{D_{11}}\tilde{\phi}\epsilon^{a+d+1} & 0 & 0\\           
0 & y_{D_{22}}\tilde{\phi}\epsilon^{a+c} & 0 \\           
0 & 0 &  y_{D_{33}}\tilde{\phi}\epsilon^{a+b} \\           
 \end{pmatrix},
 M_l =               
 \begin{pmatrix}        
y_{l_{11}}\phi\epsilon^{a+1} & 0 & 0\\           
0 & y_{l_{22}}\phi\epsilon^{a+1} & 0 \\           
0 & 0 &  y_{l_{33}}\phi \epsilon^{a+2} \\           
 \end{pmatrix}$}.
 \end{equation}
 \;
 \;
 \;
 \;
 \;
 \;
 \;

We get the effective neutrino mass matrix $M_{\nu}$ using seesaw mechanism $M_\nu =-M_DM_R^{-1}M_D^{T}$ as 
\begin{equation}
\resizebox{1 \textwidth}{!} 
 {$M_\nu =\Omega
 \begin{pmatrix}        
 \epsilon^2\tilde{\phi}^2y_{D_{11}}^2({y_{\chi_{23}}^{(2)}}^2\chi_{2}^2-m_{33}y_{\chi_{22}}^{(3)}\chi_{3}) & \epsilon\tilde{\phi}^2y_{D_{11}}y_{D_{22}}(m_{11}m_{33}-y_{\chi_{13}}^{(1)}y_{\chi_{23}}^{(2)}\chi_1\chi_2) & -\epsilon\tilde{\phi}^2y_{D_{11}}y_{D_{33}}(m_{11}y_{\chi_{23}}^{(2)}\chi_{2}-y_{\chi_{13}}^{(1)}y_{\chi_{22}}^{(3)}\chi_{1}\chi_{3})\\           
 \epsilon\tilde{\phi}^2y_{D_{11}}y_{D_{22}}(m_{11}m_{33}-y_{\chi_{13}}^{(1)}y_{\chi_{23}}^{(2)}\chi_1\chi_2) & \tilde{\phi}^2 y_{D_{22}}^2{y_{\chi_{13}}^{(1)}}^2\chi_{1}^2 & -m_{11}\tilde{\phi}^2y_{D_{22}}y_{D_{33}}Y_{\chi_{13}}^{(1)}\chi_{1} \\           
 -\epsilon\tilde{\phi}^2y_{D_{11}}y_{D_{33}}(m_{11}y_{\chi_{23}}^{(2)}\chi_{2}-y_{\chi_{13}}^{(1)}y_{\chi_{22}}^{(3)}\chi_{1}\chi_{3}) & -\tilde{\phi}^2y_{D_{22}}y_{D_{33}}Y_{\chi_{13}}^{(1)}\chi_{1}m_{11}  & \tilde{\phi}^2y_{D_{33}}^2m_{11}^2\\           
 \end{pmatrix}$},
\end{equation}
where $\Omega=\frac{\epsilon^{2a}}{m_{11}^2m_{33}-y_{\chi_{13}}^{(1)}\chi_{1}(2m_{11}y_{\chi_{23}}^{(2)}\chi_{2}-y_{\chi_{13}}^{(1)}y_{\chi_{22}}^{(3)}\chi_{1}\chi_{3})}$.\par
\;\;
\;\;
\;\;
\;\;
\;\;
\;\;

\underline{Symmetry realization for $C_{12}=0$}\par
\begin{center}
$M_R =               
 \begin{pmatrix}        
 \gamma & 0 & \zeta\\           
 0 & \eta & \upsilon \\           
 \zeta & \upsilon & \kappa \\           
 \end{pmatrix},
 \;
 M_D =               
 \begin{pmatrix}        
 x & 0 & 0\\           
 0 & y & 0 \\           
 0 & 0 & z \\           
 \end{pmatrix}$,
 \end{center}
 \begin{equation}\label{(44)}
  M_\nu =              
 \frac{1}{\Psi}\begin{pmatrix}        
 (-\upsilon^2+\eta\kappa)x^2 & \zeta\upsilon xy & -\zeta\eta xz \\           
 \zeta\upsilon xy & (-\zeta^2+\gamma\kappa)y^2 & -\gamma\upsilon yz \\           
 -\zeta\eta xz & -\gamma\upsilon yz & \gamma\eta z^2 \\           
 \end{pmatrix},
 \end{equation}
 where $\Psi=-\zeta\eta^2-\gamma\upsilon^2+\gamma\eta\kappa$.

\begin{table}[H]
\caption{Symmetry transformation for $C_{12}=0$.}
\centering
\begin{tabular}{c c c c c c}
\hline
{} & {} & Symmetry under $Z_5$ & {} & {}\\
\hline
$\nu_{R1}\rightarrow \nu_{R1}$ & $\nu_{R2}\rightarrow \omega^2\nu_{R2}$ &  $\nu_{R3}\rightarrow \omega^3\nu_{R3}$ &
$\overline{D}_{L1}\rightarrow \overline{D}_{L1}$ & $\overline{D}_{L2}\rightarrow\omega^3\overline{D}_{L2}$ & $\overline{D}_{L3}\rightarrow \omega^2\overline{D}_{L3}$ \\
\hline
$l_{R1}\rightarrow l_{R1}$ & $l_{R2}\rightarrow \omega^2 l_{R2}$ & $l_{R3}\rightarrow \omega^3 l_{R3}$ & $\chi_{1}\rightarrow \omega^2\chi_{1}$ & $\chi_2\rightarrow \omega\chi_2$ & $\chi_3\rightarrow\omega^4\chi_3$\\
\hline
$\phi\rightarrow \phi $ & {} &{} & {} & {} \\
\hline
\end{tabular}
\newline\newline

\end{table}
\underline{Symmetry realization for $C_{13}=0$}

\begin{center}
$M_R =               
 \begin{pmatrix}        
 \gamma & \xi & 0\\           
 \xi & \eta & \upsilon \\           
 0 & \upsilon & \kappa \\           
 \end{pmatrix},
 \;
 M_D =               
 \begin{pmatrix}        
 x & 0 & 0\\           
 0 & y & 0 \\           
 0 & 0 & z \\           
 \end{pmatrix}$,
 \end{center}
 \begin{equation}\label{(45)}
  M_\nu =              
 \frac{1}{\Sigma}\begin{pmatrix}        
 (-\upsilon^2+\eta\kappa)x^2 & -\xi\kappa xy & \xi\upsilon xz \\           
 -\xi\kappa xy &  \gamma\kappa y^2 & -\gamma\upsilon yz \\           
 \xi\upsilon xz & -\gamma\upsilon yz & (-\xi^2+\gamma\eta)z^2 \\           
 \end{pmatrix},
 \end{equation}
 where $\Sigma=-\gamma\upsilon^2-\xi^2\kappa + \gamma\eta\kappa$. 
\begin{table}[H]
\caption{Symmetry transformation for $C_{13}=0$.}
\centering
\begin{tabular}{c c c c c c}
\hline
{} & {} & Symmetry under $Z_5$ & {} & {}\\
\hline
$\nu_{R1}\rightarrow \omega\nu_{R1}$ & $\nu_{R2}\rightarrow \nu_{R2}$ &  $\nu_{R3}\rightarrow \omega^3\nu_{R3}$ &
$\overline{D}_{L1}\rightarrow \omega^4\overline{D}_{L1}$ & $\overline{D}_{L2}\rightarrow \overline{D}_{L2}$ & $\overline{D}_{L3}\rightarrow \omega^2\overline{D}_{L3}$ \\
\hline
$l_{R1}\rightarrow\omega l_{R1}$ & $l_{R2}\rightarrow l_{R2}$ &$l_{R3}\rightarrow \omega^3 l_{R3}$ & $\chi_{1}\rightarrow \omega^3\chi_{1}$ & $\chi_{2}\rightarrow \omega^4\chi_{2}$ & $\chi_3\rightarrow\omega^2\chi_3$\\
\hline
 $\phi\rightarrow \phi $ & {} &{} & {} & {} \\
\hline
\end{tabular}
\newline\newline

\end{table}
\underline{Symmetry realization for $C_{22}=0$}

\begin{center}
$M_R =               
 \begin{pmatrix}        
 \gamma & \xi & \zeta\\           
 \xi & 0 & \upsilon \\           
 \zeta & \upsilon & \kappa \\           
 \end{pmatrix},
 \;
 M_D =               
 \begin{pmatrix}        
 x & 0 & 0\\           
 0 & y & 0 \\           
 0 & 0 & z \\           
 \end{pmatrix}$,
 \end{center}
 \begin{equation}\label{(46)}
  M_\nu =              
 \frac{1}{\Delta}\begin{pmatrix}        
 -\upsilon^2 x^2 & (\zeta\upsilon-\xi\kappa)xy & \xi\upsilon xz\\           
 (\zeta\upsilon-\xi\kappa)xy & (-\zeta^2+\gamma\kappa)y^2 & (\xi\zeta-\gamma\upsilon)yz \\           
\xi\upsilon xz & (\xi\zeta-\gamma\upsilon)yz & -\xi^2z^2 \\           
 \end{pmatrix},
 \end{equation}
where $\Delta=2\xi\zeta\upsilon-\gamma\upsilon^2-\xi\kappa^2$. 
\begin{table}[H]
\caption{Symmetry transformation for $C_{22}=0$.}
\centering
\begin{tabular}{c c c c c c}
\hline
{} & {} & Symmetry under $Z_5$ & {} & {}\\
\hline
$\nu_{R1}\rightarrow \omega^2\nu_{R1}$ & $\nu_{R2}\rightarrow \omega^3\nu_{R2}$ &  $\nu_{R3}\rightarrow \omega\nu_{R3}$ &
$\overline{D}_{L1}\rightarrow \omega^3\overline{D}_{L1}$ & $\overline{D}_{L2}\rightarrow \omega^2\overline{D}_{L2}$ & $\overline{D}_{L3}\rightarrow \omega^4\overline{D}_{L3}$ \\
\hline
$l_{R1}\rightarrow\omega^2 l_{R1}$ & $l_{R2}\rightarrow \omega^3 l_{R2}$ & $l_{R3}\rightarrow \omega l_{R3}$ & $\chi_{1}\rightarrow \omega\chi_{1}$ & $\chi_2\rightarrow \omega^2\chi_2$ & $\chi_3\rightarrow\omega^3\chi_3$\\
\hline
$\phi\rightarrow \phi $ & {} &{} & {} & {} \\
\hline
\end{tabular}
\newline\newline

\end{table} 
\underline{Symmetry realization for $C_{33}=0$}

\begin{center}
$M_R =               
 \begin{pmatrix}        
 \gamma & \xi & \zeta\\           
 \xi & \eta & \upsilon \\           
 \zeta & \upsilon & 0 \\           
 \end{pmatrix},
 \;
 M_D =               
 \begin{pmatrix}        
 x & 0 & 0\\           
 0 & y & 0 \\           
 0 & 0 & z \\           
 \end{pmatrix},$
 \end{center}
 \begin{equation}\label{(47)}
  M_\nu =              
 \frac{1}{\Pi}\begin{pmatrix}        
 -\upsilon^2 x^2 & \zeta\upsilon xy & (-\zeta\eta+\xi\upsilon)xz\\           
  \zeta\upsilon xy &  -\zeta^2y^2 & (\xi\zeta-\gamma\upsilon)yz \\           
 (-\zeta\eta+\xi\upsilon)xz & (\xi\zeta-\gamma\upsilon)yz & (-\xi^2+\gamma\eta)z^2 \\           
 \end{pmatrix},
 \end{equation}
where $\Pi=-\zeta^2\eta+2\xi\zeta\upsilon-\gamma\upsilon^2$. 
\begin{table}[H]
\caption{Symmetry transformation for $C_{33}=0$.}
\centering
\begin{tabular}{c c c c c c}
\hline
{} & {} & Symmetry under $Z_5$ & {} & {}\\
\hline
$\nu_{R1}\rightarrow \nu_{R1}$ & $\nu_{R2}\rightarrow \omega\nu_{R2}$ &  $\nu_{R3}\rightarrow \omega^2\nu_{R3}$ &
$\overline{D}_{L1}\rightarrow \overline{D}_{L1}$ & $\overline{D}_{L2}\rightarrow \omega^4\overline{D}_{L2}$ & $\overline{D}_{L3}\rightarrow \omega^3\overline{D}_{L3}$ \\
\hline
$l_{R1}\rightarrow l_{R1}$ & $l_{R2}\rightarrow \omega l_{R2}$ & $l_{R3}\rightarrow \omega^2 l_{R3}$ & $\chi_{1}\rightarrow \omega^4 \chi_{1}$ & $\chi_2\rightarrow \omega^3\chi_2$ & $\chi_3\rightarrow\omega^2\chi_3$\\
\hline
$\phi\rightarrow \phi $ & {} &{} & {} & {} \\
\hline
\end{tabular}
\newline\newline

\end{table}
For all the viable cases we obtain
\begin{equation}
 M_\nu \sim \epsilon^{2a}
 \begin{pmatrix}        
 \epsilon^2 & \epsilon & \epsilon\\           
 \epsilon & 1 & 1 \\           
 \epsilon & 1 & 1 \\           
 \end{pmatrix}.
\end{equation}

The texture of $M_{\nu}$ indicates normal hierarchy with $\mu-\tau$ symmetry i.e, $\theta_{13}=0$ and the maximal atmospheric mixing $\theta_{23}=\frac{\pi}{4}$. To achieve experimentally viable textures, broken $\mu-\tau$ symmetry and deviation from maximal atmospheric mixing can be done by appropriate pertubation in the neutrino mass matrix.
\par

\section{RESULTS AND DISCUSSION}
In this work, we have carried out a phenomenological texture study of the Majorana neutrino mass matrices with the ansatzes of one vanishing minor and the zero sum of the eigenvalues with the CP phases. One of the two simultaneous constraint equations consists of the cross term of the variables, so we had option of four solution pairs of the equations. Interestingly the solution pairs have interplay in various possible textures under study. The systematic numerical analysis has been carried out with the latest $3\sigma$ neutrino oscillation data. Although the current neutrino oscillation data shed some light on the range of the Dirac CP phase $\delta$, but the Majorana CP phases $\alpha$ and $\beta$ are still completely unexplored. As the prime objective of this work to step in such unknown terrain of neutrinos, we have strategized to find out the phenomenologically allowed values of the Majorana CP phases $\alpha$ and $\beta$ for different viable textures. We have also explored the neutrinoless double beta decay rate, $|m_{ee}|$ and the strength of CP violation, $J_{cp}$ for all viable textures. The ranges of $\alpha$, $\beta$, $\delta$, $|m_{ee}|$ and $J_{cp}$ in our study have been summarized in Table III. \par

To understand the origin of zeros in fermion mass matrices, we have implemented $Z_5$ flavor symmetry group. Again to get the information of hierarchy of the viable textures, additionally the FN mechanism was augmented. The symmetry realization is an important work for realistic model building.\par

Now we summarize our observations of this texture study as follows:\\
\begin{itemize}
\item[a.] The viability of the textures was checked on the basis of the values of $\delta$ within the values of the ratio of the mass squared difference $R_{\nu}$ both at $3\sigma$ level. In this context, the textures $C_{11}=0$, $C_{13}=0$, $C_{22}=0$ and $C_{33}=0$ have been found viable for normal hierarchy only and the case $C_{12}=0$ has been found viable for both normal and inverted hierarchies. Again the case $C_{23}=0$ is completely ruled out. Interestingly the solution pair $(X_{+}, Y_{-})$ from our ansatzes supports all the cases except the case $C_{13}=0$. The solution pairs $(X_{+}, Y_{+})$ and $(X_{-}, Y_{-})$ support the cases $C_{12}=0$ and $C_{13}=0$ for normal hierarchy. Further the solution pair $(X_{-}, Y_{+})$  supports $C_{12}=0$ for inverted hierarchy and $C_{22}=0$ for normal hierarchy. The interplay of the solution pairs exists in the results.

\item[b.] The Majorana phase $\alpha$ for the textures $C_{12}=0$ and $C_{13}=0$ is vanishingly small and the range is highly constrained. 

\item[c.] For all the viable textures, the atmospheric mixing angle $\theta_{23}$ lies in the range ($40^{\degree}$, $45^{\degree}$). Thus the phenomenology of these textures favors the first quadrant for atmospheric mixing.
 
\item[d.] For all the cases both the neutrinoless double beta decay rate, $|m_{ee}|$ and the strength of the Dirac CP violation, $J_{CP}$ remain within the experimental bounds.

\item[e.] Symmetry realization of all the viable textures has been done under the discrete symmetry group $Z_{5}$. Additionally FN mechanism has been augmented to check the hierarchy of the textures. We have found that all the cases favour normal hierarchy of neutrino mass pattern. \par
\end{itemize}
Finally, we expect that our numerical results of the Dirac and Majorana CP phases may be verified in the future neutrino experiments designed for the purpose.

\section*{ACKNOWLEDGEMENT}
One of the authors SD has acknowledged the funding support for this work from the Department of Science and Technology (DST), India (Grant DST/INSPIRE Fellowship/2018/IF180713) under the scheme of INSPIRE fellowship programme.

\bibliographystyle{ieeetr}
\bibliography{VanMinor_aRxiv}

\end{document}